\newcommand\ee{e E} %{e Q \frac{B}{A}}

\newcommand{\rr}{\mathbb{R}}

\newcommand{\oom}{\pi\omega}

\newcommand\beq{\begin{equation}}
\newcommand\eeq{\end{equation}}
\newcommand\beqnl{\begin{eqnarray}}
\newcommand\beqna{\begin{eqnarray*}}
\newcommand\eeqna{\end{eqnarray*}}
\newcommand\eeqnl{\end{eqnarray}}
\newcommand{\Ree}{Re}
\newcommand{\trace}{Tr}
\newcommand{\Imm}{Im}

\newcommand\NN{{\mathbb N}}
\newcommand\RR{{\mathbb R}}

\documentclass[12pt]{iopart}
\usepackage{amsfonts}
\usepackage{setstack}
\usepackage{amsthm}
\usepackage{amsbsy}
\usepackage{amssymb}

\begin{document}

\date{\today}

\title[Quantum instability on charged
 Nariai and ultracold black hole manifolds]{Quantum instability for charged scalar particles on charged
 Nariai and ultracold black hole manifolds}

\author{F. Belgiorno\footnote{E-mail address: belgiorno@mi.infn.it}}
\address{Dipartimento di Matematica, % and Dipartimento di Fisica,
Universit\`a di Milano, 20133
Milano, Italy}
\author{S. L. Cacciatori\footnote{E-mail address: sergio.cacciatori@uninsubria.it}}
\address{Dipartimento di Matematica e Fisica, Universit\`a dell'Insubria, 22100
Como, Italy, and\\
I.N.F.N., sezione di Milano, Italy}
\author{F. Dalla Piazza\footnote{E-mail address: f.dallapiazza@uninsubria.it}}
\address{Dipartimento di Matematica e Fisica, Universit\`a dell'Insubria, 22100
Como, Italy, and\\
I.N.F.N., sezione di Milano, Italy}

\begin{abstract}

We analyze in detail the quantum instability which characterizes charged scalar
field on three special de Sitter charged black hole backgrounds. In particular, we
compute exactly the imaginary part of the effective action for scalar charged
fields on the ultracold I, ultracold II and Nariai charged black hole backgrounds.
Both the transmission coefficient approach and the $\zeta$-function approach are
exploited.
Thermal effects on this quantum instability are also taken into account in presence
of a non-zero black hole temperature (ultracold I and Nariai).

\end{abstract}

%\pacs{04.62.+v, 04.70.Dy}
\maketitle

\section{Introduction}
\label{intro}

It is well-known that quantum effects lead to the loss of charge by charged black
holes \cite{gibbons,khriplovich,damo,deruelle},
and that this phenomenon on one hand is independent on the fact that
there can be a contextual evaporation phenomenon (i.e. also extremal black holes,
with zero temperature, are involved in this spontaneous loss of charge), on the
other hand its being related to the Schwinger instability of vacuum in presence
of a constant electric field has been pointed out. The latter topic can be
brought back to the more general class of phenomena of quantum instability
in presence of an external field, which still has in the Schwinger calculation
the most relevant and important contribution \cite{euler-heisenberg,schwinger}.
See also \cite{gitman,kimpage}.\\
In previous studies devoted to this topic in the case of black hole backgrounds, we
mainly focused our attention to black holes of the Kerr-Newman family \cite{belmart},
also in presence of a cosmological constant \cite{belcaccia-ads,belcaccia-rnds}.
In the latter case, we were able to
perform an exact calculation for charged Dirac fields in four dimensions in three special cases
\cite{belcadalla}:
the ultracold I, ultracold II and Nariai charged black hole backgrounds. The calculations
were performed both in the so-called transmission coefficient approach and in the zeta-function
approach, obtaining so a double check for our calculations.\\
Herein, we complete our analysis performed in \cite{belcadalla} by taking into account the case of
charged scalar fields, and provide an exact calculation for their instability on the given black hole
backgrounds. We recall that in similar calculations one is far from being able to
reach exact results (e.g. in the Reissner-Nordstr\"{o}m case only a WKB approximation
is available).  We also point out that,
contrarily to what one could naively expect, the scalar field analysis
presents to some extent more difficulties than the analysis for the Dirac field, because
of some mathematical subtleties occurring in the scalar case: we limit ourselves to mention the (open)
problem of a rigorous mathematical setting for the Klein-Gordon equation minimally coupled with
an external electrostatic potential in presence of event horizon(s), requiring an analysis involving the
so-called Krein spaces in place of the more standard Hilbert spaces occurring in the analysis of
the Dirac equation. Still, we can perform with some ingenuity zeta-function calculations
and show that the imaginary part of the effective action coincides with the one calculated
by means of the transmission coefficient approach. Another peculiar behavior emerges in the scalar
field case when one takes into account the behavior of the field in the ultracold I case:
a bad behavior at infinity occurs for the wave function, but an analysis in terms of fluxes
allows to determine the transmission coefficient. Moreover, in the Nariai case,
the scalar nature of the particle is at the root of the possibility to obtain a change of sign
in a quantity $\Delta$ (cf. eqn. (\ref{delta})) due to the presence of a term $-\frac{1}{4}$
which is instead missing in the analogous quantity for the Dirac case (see \cite{belcadalla}).
This may cause a change in the behavior of the imaginary part of the effective action,
as we shall see.\\
\indent In the cases ultracold I and Nariai, which are involved with a non-zero background
temperature, thermal effects on the quantum instability are also considered.\\

The plan of the paper is the following. In section \ref{vactherm} we sum up some
aspect of the transmission coefficient approach which are relevant for our paper,
and then extend our analysis \cite{belcadalla} concerning instability of thermal state
induced by the pair-creation effect to scalar fields. In sections \ref{ultracold},
\ref{sec-ucI} and \ref{sec-nariai} we take into account the cases of ultracold II,
ultracold I and Nariai charged black hole backgrounds respectively. In section
\ref{conclusion} conclusions are drawn.

\section{Vacuum instability and Thermal state instability}
\label{vactherm}

We discuss in the following some aspect of the problem of vacuum instability and of thermal
state instability induced by it. The Dirac case was discussed in \cite{belcadalla}.

\subsection{Vacuum instability}

For completeness, we summarize some aspect of the transmission coefficient approach in the case of
scalar fields, following \cite{damo,nikishov,naroz}.
We are mainly interested in the probability of persistence of the vacuum.
Let us introduce, for a diagonal scattering process \cite{damo}
\beq
n_i^{IN}= R_i n^{OUT}_i + T_i p^{OUT}_i,
\eeq
where $n_i$ stays for a negative energy mode and $p_i$ for a positive energy one.
$T_i$ is the transmission coefficient and $R_i$ is the reflection one.
Moreover, as in \cite{damo}, we define
\beq
\eta_i:=|T_i|^2.
\eeq
Then, it is possible to show that for bosons one gets
\beq
|R_i|^2 = 1+\eta_i,
\eeq
which accounts for the well-known superradiance phenomenon.\\
The persistence of the vacuum is given by \cite{damo}
\beq
P_0 = \prod_i p_{i,0} = e^{-2 \Imm W},
\eeq
where $p_{i,0}$ is the probability to have zero pair in channel $i$, and then
\beq
2 \Imm W = \sum_i  \log (1+\eta_i) = \sum_i \sum_{k=1}^{\infty} (-)^{k+1} \frac{1}{k} \eta_i^k.
\eeq

\subsection{Thermal state instability}

We have to take into account that, in the case of the ultracold I manifold and also in the
Nariai charged case, there exists an intrinsic thermality of the background manifold
which is associated with the presence of non-degenerate horizons. As a consequence, the real
quantum state to be considered is not the vacuum state in a traditional sense (i.e. absence of
particles), but the thermal state associated with the aforementioned temperature (we recall that
we are dealing with special manifolds endowed with a single temperature even if two different
non-degenerate event horizons are involved). As a consequence, we construct the Hartle-Hawking
state for our thermal geometries, by adopting the same attitude as in \cite{belcadalla}.\\
We point out that the following construction holds true in general, even if we are interested
in it for our specific analysis. We adopt the thermofield dynamics formalism, and define a
thermal state $|0(\beta)>$ characterized by an inverse temperature $\beta$. This state
is annihilated by suitable operators $a_l (\beta),\tilde{a}_l(\beta)$,
$b_l(\beta),\tilde{b}_l(\beta)$ (and conjugated ones) which are labeled by a complete set of
quantum numbers $l$ and is related to ``standard''
annihilation-creation operators $a_l,\tilde{a}_l,b_l,\tilde{b}_l$ (and conjugated ones)
via a formally unitary transformation:
\beqnl
a_l &=& c^+_l (\beta) a_l (\beta) + s^+_l (\tilde{a}_l)^{\dagger} (\beta),\\
b_l &=& c^-_l (\beta) b_l (\beta) + s^-_l (\tilde{b}_l)^{\dagger} (\beta),
\eeqnl
and analogues for hermitian conjugates, with
\beqnl
s^+_l &=& \frac{1}{\sqrt{e^{\beta (\omega_l -\phi^+)}-1}}\\
s^-_l &=& \frac{1}{\sqrt{e^{\beta (|\omega_l| +\phi^-)}-1}},
\eeqnl
where $\phi^+,\phi^-$ stay for chemical potentials for particles and antiparticles respectively.
Moreover, it holds
\beqnl
(c^+_l)^2-(s^+_l)^2 &=& 1,\\
(c^-_l)^2-(s^-_l)^2 &=& 1.
\eeqnl
We also introduce standard Bogoliubov relations between ``in'' and ``out'' operators
as follows:
\beqnl
a_l^{out} &=& \rho_l\; a_l^{in}  + T^{\ast}_l (b_l^{in})^{\dagger} ,\\
b_l^{out} &=& \rho_l\; b_l^{in}  + T^{\ast}_l (a_l^{in})^{\dagger},
\eeqnl
where $|\rho_l|^2-|T_l|^2 =1$. Compare also \cite{kimleeyoon}. Note that we limited
ourselves to consider diagonal transformations, as it is the interesting case for our
considerations.\\

In order to check how thermal effects affect instability of quantum fields, we consider, in
place of the usual
$<0\; in|(a_l^{out})^{\dagger} a_l^{out} |0\; in>$, which gives the number
of out-particles on the in-vacuum, the following quantity:
\beq
N_{out}^+ := <0(\beta)\; in|(a_l^{out})^{\dagger} a_l^{out} |0(\beta)\; in>,
\eeq
and check if deviations from pure thermality appears in the distribution.
Equivalently, as in \cite{kimleeyoon}, we can define
\beq
\bar{N}_{out}^+ := <0(\beta)\; in|(a_l^{out})^{\dagger} a_l^{out}-
(a_l^{in})^{\dagger} a_l^{in}|0(\beta)\; in>,
\eeq
which just signals us the deviation part (or it is zero).\\
It is easily shown that
\beq
\bar{N}_{out}^+ = |T_l|^2 \left[ (s^+_l)^2 + (s^-_l)^2 + 1 \right],
\eeq
where $|T_l|^2$ is the transmission coefficient.
When $\phi^+=\phi^-=\phi$, as in the case of our black hole background, we obtain
\beq
\bar{N}_{out}^+ = |T_l|^2 \frac 12 \left[ \coth \left( \frac{\beta (\omega_l -\phi)}{2} \right) +
\coth \left( \frac{\beta (|\omega_l| +\phi)}{2} \right) \right],
\label{thermal-eff}
\eeq
which is easily realized to coincide with the result displayed for the boson case
in \cite{kimleeyoon} when $\phi=0$, and matches the results in \cite{belcadalla}
for the Dirac case. Note that (\ref{thermal-eff}) can be used also for the
Reissner-Nordstr\"{o}m case, where the coefficient $|T_l|^2$ is known only in
the WKB approximation \cite{gibbons,khriplovich}.

\section{Ultracold II case}
\label{ultracold}

The ultracold II metric is obtained from the Reissner-Nordstr\"om-de Sitter one
in the limit of coincidence of the Cauchy horizon, of the black hole event horizon
and of the cosmological event horizon. See \cite{romans,mann}. In particular,
the metric we are interested in is
\beq
ds^2 = -d t^2 + dy^2+ \frac{1}{2\Lambda} (d\theta^2+
\sin^2 (\theta) d\phi^2),
\label{ultracold-II}
\eeq
with $y\in \RR$ and $t\in \RR$.
The electromagnetic field strength is $F=-\sqrt{\Lambda}  dt\wedge dy$, and we
can choose $A_0 = \sqrt{\Lambda} y$ and $A_j=0$, $j=1,2,3$. It is also useful to
define $E := \sqrt{\Lambda}$, which represents the modulus of the electrostatic
field on the given manifold. We note that it is uniform, and then one expects naively
to retrieve at least some features of Schwinger's result, as in the Dirac case
\cite{belcadalla}.

\subsection{The transmission coefficient approach}

Let us consider the Klein-Gordon equation in the given manifold
\beq
[-(-i \partial_t+e E y)^2 -\partial_y^2 -2 \Lambda \nabla^2_{\Omega} + \mu^2]\phi = 0,
\eeq
where $\mu$ and $e$ are the mass and the charge of the scalar particle. We assume $\ee>0$
for definiteness.
In agreement with the possibility to perform variable separation, let us set
\beq
\phi (t,y,\Omega) = e^{-i \omega t} Y_{l m} (\Omega) \psi (y);
\eeq
then one obtains the following equation for $\psi$:
\beq
\frac{d^2 \psi}{dy^2} (y) = (\mu_l^2 -(\omega + e E y)^2 ) \psi (y),
\label{eq-psi}
\eeq
where $\mu_l^2=2\Lambda l(l+1)+\mu^2$.
By defining (cf. \cite{damo})
\beqnl
\xi &=& \frac{1}{\sqrt{e E}}(\omega + e E y),\cr
\lambda &=& \frac{1}{e E} \mu_l^2,\cr
k &=& -\frac{1}{2}- i \frac{\lambda}{2},\cr
u &=& \sqrt{2} e^{-i \frac{\pi}{4}} \xi,
\eeqnl
one obtains
\beq
\frac{d^2 \psi}{d\xi^2} (\xi) = (\lambda -\xi^2 ) \psi (\xi),
\eeq
whose solution is
\beq
\psi = D_n (u),
\eeq
which is a parabolic cylinder function. The calculation is completely analogous to the
one performed in \cite{damo}, and as in \cite{damo} one can easily show that the
transmission coefficient satisfies
\beq
|T_l|^2 = e^{-\pi \lambda} = e^{ -\pi \frac{\mu_l^2}{e E} }.
\eeq
The latter expression coincides with the WKB approximation for the same coefficient \cite{belcaccia-rnds}
(that calculation is for Dirac particles, but it is easy to realize that for scalar particles
the result is the same, apart for the obvious replacement $k^2\mapsto l(l+1)$).
This means that the WKB approximation is actually exact for the given case.
We have the exact transmission coefficient. As in \cite{damo,khriplovich} we can determine the
degeneracy factor, and one obtains
\beq
W=\frac{e E S}{2\pi} \sum_{l=0}^{\infty} (2 l +1)
\log (1+ e^{-\frac{\pi \mu_l^2}{e E} } ),
\label{ucII-w}
\eeq
where $S$ is the spacetime volume of the $(t,y)$-part of the manifold.

\subsection{The $\zeta$-function approach}

We can use the $\zeta$-function regularization to compute the effective action.
The spectral zeta function for the Euclidean Klein-Gordon equation is given by
\begin{eqnarray}
\zeta (s)=\frac 1{\Gamma(s)}\int_0^\infty z^{s-1} K(z) dz,
\end{eqnarray}
with kernel
\begin{eqnarray}
K (z)={\mathrm{Tr}} e^{-[-(\partial_{\tau}+i e E y)^2 -\partial_y^2 -2 \Lambda \nabla^2_{\Omega} + \mu^2] z},
\end{eqnarray}
where $\tau$ stays for the Euclidean time.
To compute the trace, note that the operator $-2 \Lambda \nabla^2_{\Omega} + \mu^2$ commutes with the Klein-Gordon operator so that
it contributes with the eigenvalues $\mu^2_{l}=2\Lambda l(l+1) +\mu^2$ with degeneration $(2l+1)$.
Next, noting that the operator
$\hat p:=-i\partial_{\tau}$ commutes with $\hat A:=-(\partial_t+i e E y)^2 -\partial_y^2$ we can restrict on the eigenspaces having eigenvalue $\omega$
for $\hat p$. Thus, $\hat A=(\omega+e E y)^2 -\partial_y^2$ which describe a harmonic oscillator with eigenvalues $eE (2n+1)$. Independence
on $\omega$ shows that such eigenvalues are degenerate so that if $D$ is the degeneration we can write
\begin{eqnarray}
K(z)=\sum_{l=0}^\infty \sum_{n=0}^\infty l(l+1) D e^{-[eE (2n+1)+\mu_l^2] z}=
\sum_{l=0}^\infty l(l+1) D \frac {e^{[\ee-\mu_l^2] z}}{e^{2 \ee z}-1}.
\end{eqnarray}
We can determine the degeneration factor as done in \cite{blissepf}. We then obtain $D= \frac {eE}{2\pi} S$, where
$S$ is the spacetime volume of the $(t,y)$-part of the manifold. \\
The Euclidean action is
$$
S_E= -\zeta' (0),
$$
and one finds
$$
W=\Imm S_L,
$$
where $S_L$ is the Lorentzian action, in our case obtained by $E\rightarrow iE$. Explicitly
\begin{eqnarray}
\zeta (s) =\sum_{l=0}^\infty \frac {eE}{2\pi} S (2l+1)
\left(\frac{\gamma^2}{2eE}\right)^s \zeta_H (s; \frac {\mu^2_l}{2eE} +\frac 12 )
\end{eqnarray}
where $\zeta (s;a)$ is the Hurwitz zeta function and $\gamma$ is a renormalization scale, henceforth
put equal to 1 (also in the ultracold I and Nariai case). We also put
\begin{equation}
W=\sum_{l=0}^\infty W_l.
\label{wl-sum}
\end{equation}
After Lorentzian continuation we find
$$
W_l=\frac {eE S}{2\pi} (2l+1) \Ree \left[+i \frac {\mu^2_l}{2eE}  \log (2ieE)+\frac 12 \log {2\pi} -\zeta'_H (0; -i \frac {\mu_l^2}{2eE}+\frac 12)
 \right].
$$
One can notice that:
$$
\Ree \left[i \frac {\mu^2_l}{2eE} i \frac{\pi}{2}+\frac 12 \log {2\pi} -\log \Gamma(\frac 12 - i
\frac {\mu_l^2}{2eE}) \right]=
 \frac 12 \log (1+e^{-\pi \frac {\mu^2_l}{eE}}).
$$
Thus, the final expression for the imaginary part of the Lorentzian action is:
\beq
W_l=\frac {eE S}{2\pi} (2l+1)\frac 12 \log (1+e^{-\pi \frac {\mu^2_l}{eE}}),
\eeq
which coincides with (\ref{ucII-w}).

%%%%%%%%%%%%%%%%%%%%%%%%%%%%%%%%%%%%%%%%%%%%%%%%%%%%%%%%%%%%%%%%%%%%%%%%%%%%%%%%%%%%%%%%%%%%%%%%%%%%%%%%%%

\section{The ultracold I case}
\label{sec-ucI}

A second extremal limit of the Nariai background is given by the type I ultracold solution
when  $r_-=r_+=r_c$.
The metric is \cite{mann}
\beq
ds^2 = -\chi^2 d \psi^2 + d\chi^2+ \frac{1}{2\Lambda} (d\theta^2+
\sin^2 (\theta) d\phi^2),
\label{ultracold-I}
\eeq
with $\chi\in (0,\infty)$ and $\psi\in \RR$, and the electromagnetic field strength is
$F=\sqrt{\Lambda} \chi d\chi\wedge d\psi$. The spacetime presents the structure of a
2D Rindler manifold times a two dimensional sphere (with a constant warping factor).
One gets $\Gamma_{01}^0 = \frac{1}{\chi},
\Gamma_{00}^1 = \chi, \Gamma_{33}^2 = -\sin (\theta) \cos (\theta),
\Gamma_{23}^3 = \cot (\theta)$.
We can choose $A_0 = \frac{\sqrt{\Lambda}}{2} \chi^2$ and $A_j=0$, $j=1,2,3$ as potential.
This case is a little bit more tricky than the ultracold II and we will adopt a different strategy to define the transmission and reflection coefficients. However, we will again be able to compare this approach with the zeta function method and the two results are the same. The expression of the effective action for a scalar field in this background fits our previous result for the Dirac case \cite{belcadalla}.

\subsection{The transmission coefficient approach}

In order to compute the wave functions for a scalar field in this background we search the solutions of the Klein-Gordon equation, for the variable $\chi$, that for this metric is:
\begin{equation}
\left[ -\frac 1{\chi^2} (\omega+eE\frac {\chi^2}2)^2-\frac 1\chi \partial_\chi (\chi\partial_\chi)+\mu_l^2 \right]\Psi=0,
\end{equation}
where to perform variable separation we pose $\phi(\tau,\chi,\Omega)=e^{-i\omega\tau} Y_{lm}(\Omega)\Psi(\chi)$.
Making the change of variable $t=\frac{\chi^2}2$:
\beq
\left[ (t\partial_t)^2+\frac 14(\omega+eEt)^2-\frac t2 \mu_l^2  \right]\Psi=0.
\eeq
Next, we set, as usual, the wave function in the factorized form:
\beq
\Psi(\chi(t))=e^{\frac i2 eEt}t^{i\frac \omega2} F(t),
\eeq
so we obtain the following confluent hypergeometric equation or Kummer's equation:
\beq
t\partial_t^2 F +(1+i\omega+ieEt)\partial_t F -\frac 12 (\mu_l^2-ieE)F=0,
\eeq
whose general solution is
$$
F(t)=\alpha \Phi (\frac 12-\frac {\mu_l^2}{2ieE}; 1+i\omega;-ieEt)+\beta t^{-i\omega} \Phi (\frac 12-\frac {\mu_l^2}{2ieE}-i\omega; 1-i\omega;-ieEt),
$$
where $\Phi(a;c;z)$ is the usual Kummer function (or 1-st kind confluent hypergeometric function).
Then, the general solution is
\beqnl
\Psi(\chi(t))=&&e^{\frac i2 eEt}\left[\alpha t^{\frac i2 \omega} \Phi (\frac 12-\frac {\mu_l^2}{2ieE}; 1+i\omega;-ieEt)\right.\cr
&&\left.\hphantom{.........}+ \beta t^{-\frac i2 \omega}
\Phi (\frac 12-\frac {\mu_l^2}{2ieE}-i\omega; 1-i\omega;-ieEt)\right].
\eeqnl
The asymptotic behavior of the wave function can be determined using for $|z|\approx \infty$:
\beq
\Phi(a;c;z)\approx \frac {\Gamma(c)}{\Gamma(c-a)} (-z)^{-a} (1+O(1/z)) +\frac {\Gamma(c)}{\Gamma(a)} e^z z^{-(c-a)} (1+O(1/z)).
\eeq
In the ultracold I background the electric field vanishes in $t=0$ and it grows indefinitely for $t\approx\infty$. Thus, for large $t$ a charged particle is subjected to an increasing force and it is accelerated toward infinity. For this reason the particle, for $t\approx\infty$, does not behave as a free particle. The asymptotic behavior of the Kummer function reflects these physical considerations: the presence in the asymptotic expansion of the wave functions of terms proportional to $t^{-1/2}$ shows that the behavior of the particle is far to the one of a free particle. Thus, one can not define the transmission or reflection coefficients in the usual way.
However, we can define them using a slightly different strategy. First we compute the Klein-Gordon conserved current. Then, in the region $t\approx 0$ one expects to find a flux of matter coming from $t\approx 0$ and also a reflected one from large $t$. Instead, in the region $t\approx\infty$, due to the electric field, one expects just the presence of a transmitted flux, the one started from small $t$ region, and no reflected one. These considerations allow, in a quite straightforward way, to define the transmission coefficient as the ratio between  the transmitted flux at $t\approx\infty$ and the incoming flux at $t\approx 0$ and the reflection coefficient as the ratio between the reflected flux and the incoming flux at $t\approx 0$. Let $v=\frac {\partial}{\partial \psi}$ the $4$-velocity of the static coordinate observer, $j$ the conserved current and
$d\Sigma^{\mu\nu}$ the surface element through which we would compute the flux. Then, the associated infinitesimal flux
is $\Phi_\Sigma (j)=\frac {\sqrt g}2 \epsilon_{\mu\nu\rho\sigma} j^\rho v^\sigma d\Sigma^{\mu\nu}$. As we are interested to the flux in the
$x$ direction ($\chi=e^{2x}$), $d\Sigma^{\mu\nu}=[(\delta^\mu_\theta \delta^\nu_\phi-\delta^\nu_\theta \delta^\mu_\phi)/2] d^2 b$, where $d^2 b$ is an infinitesimal  surface element,  and then
$$
\Phi_\Sigma (j)=\sqrt g j^x d^2 b=\frac 1{2\Lambda} j_x d^2 b.
$$
Thus, dropping the unessential factor $1/2\Lambda$,
we can define the transmission and reflection coefficients by looking at the covariant current only.

The covariant components of the Klein-Gordon conserved current are:
$j_\mu=%-i\Psi^*\overleftrightarrow{D}_\mu\Psi=
-\frac{i}{2}\left[\Psi^*D_\mu\Psi-(D_\mu\Psi^*)\Psi\right]$. As we will compute it for the two asymptotic regions $t\approx 0$ and $t\approx \infty$, we need the expansion of the wave functions for small and large $t$.
For $t\approx 0$ we obtain:
$$
\Psi(\chi(t))\approx\alpha e^{\frac{i}{2}eEt}t^{\frac{i}{2}\omega}+\beta e^{\frac{i}{2}eEt}t^{-\frac{i}{2}\omega},
$$
making the change of variable $t=\frac{1}{2}e^{2x}$:
$$
\Psi(\chi)\approx a e^{\frac{1}{4}eEe^{2x}+i\omega x}+b e^{\frac{1}{4}eEe^{2x}-i\omega x},
$$
and restoring the time dependence $\psi$:
$$
\Psi(\chi)\approx a e^{\frac{1}{4}eEe^{2x}+i\omega x-i\omega\psi}+b e^{\frac{1}{4}eEe^{2x}-i\omega x-i\omega\psi},
$$
with $a=\alpha (\frac12)^{\frac i2\omega}$ and $b=\beta (\frac12)^{-\frac i2\omega}$.
Finally we obtain the following expression for $x$-component of the conserved current:
\beq
j_x \approx \omega\left(|a|^2-|b|^2 \right).
%=-i \left[\frac{ieE}{2}e^{2x}\left(|a|^2+|b|^2\right)+i\omega\left(|a|^2-|b|^2 \right)+i\Imm\left(a^*be^{-2i\omega x+2x}(ieE)\right)\right]
\eeq
For $t\approx \infty$, using the expansion of the Kummer function and making the change of variable as before, the asymptotic behavior of  $\Psi(\chi(t))$ is:
\begin{eqnarray*}
\Psi(\chi(t))&&\approx c_1 e^{i\left(2k_1x+\frac{k_2}{2}e^{2x}\right)-x}+c_2e^{-i \left(2k_1x+\frac{k_2}{2}e^{2x}\right)-x}\\
&&\approx c_1e^{i\frac{eE}{4}e^{2x}-x}+c_2e^{-i\frac{eE}{4}e^{2x}-x},
\end{eqnarray*}
with $k_1=\frac \omega 2-\frac{\mu_l^2}{2eE}$, $k_2=\frac{eE}{2}$ and
\begin{eqnarray*}
c_1&=&\alpha \frac{\Gamma(1+i\omega)}{\Gamma\left(\frac 12+i(\omega-\frac{\mu_l^2}{2eE} )\right)}(ieE)^{-\frac 12 -i\frac{\mu_l^2}{2eE}}2^{-i(\frac \omega 2-\frac{\mu_l^2}{2eE})+\frac 12}\cr
&&+\beta \frac{\Gamma(1-i\omega)}{\Gamma\left(\frac 12-i\frac{\mu_l^2}{2eE}\right)}(ieE)^{-\frac 12 -\frac{i \mu_l^2}{2eE}+i\omega}2^{-i(\frac \omega 2-\frac{\mu_l^2}{2eE})+\frac 12} \\
c_2&=&\alpha \frac{\Gamma(1+i\omega)}{\Gamma\left(\frac 12+i\frac{\mu_l^2}{2eE}\right)}(-ieE)^{-\frac 12 +i\frac{\mu_l^2}{2eE}-i\omega}2^{i(\frac \omega 2-\frac{\mu_l^2}{2eE})+\frac 12}\cr
&&+\beta \frac{\Gamma(1-i\omega)}{\Gamma\left(\frac 12-i(\omega-\frac{\mu_l^2}{2eE})\right)}(-ieE)^{-\frac 12 +i\frac{\mu_l^2}{2eE}}2^{i(\frac \omega 2-\frac{\mu_l^2}{2eE})+\frac 12}.
\end{eqnarray*}
Restoring the $\psi$ dependence: $\Psi\approx c_1e^{i\frac{eE}{4}e^{2x}-x-i\omega\psi}+c_2e^{-i\frac{eE}{4}e^{2x}-x-i\omega\psi}$.
The $x$-component of the conserved current is:
\beq
j_x=\frac{eE}{2}(|c_1|^2-|c_2|^2).
\eeq
Thus, the transmission and reflection coefficients are (we are considering the crossing-level region that appears for $\omega<0$):
\begin{eqnarray*}
|T_l|^2&=&-\frac{eE}{2}\frac{|c_1|^2}{\omega |a|^2} \\
|R_l|^2&=&\frac{|b|^2}{|a|^2}.
\end{eqnarray*}
As explained before, to avoid particles coming from $t\approx\infty$, we impose the condition $c_2=0$ and we obtain:
\begin{eqnarray*}
\beta&=&-\alpha\frac{\Gamma(1+i\omega)}{\Gamma(1-i\omega)}\frac{\Gamma\left(\frac 12-i(\omega-\frac{\mu_l^2}{2eE})\right)}{\left(\frac 12+i\frac{\mu_l^2}{2eE}\right)}(-ieE)^{-i\omega} \\
c_1&=&\alpha (ieE)^{-\frac 12 -i\frac{\mu_l^2}{2eE}}2^{-i(\frac \omega 2-\frac{\mu_l^2}{2eE})+\frac 12}\left[\frac{\Gamma(1+i\omega)}{\Gamma(\frac 12 +i(\omega -\frac{\mu_l^2}{2eE}))}\right.\cr
&&\left. -\frac{\Gamma(1+i\omega)\Gamma(\frac 12 -i(\omega -\frac{\mu_l^2}{2eE})}{|\Gamma(\frac 12-i\frac{\mu_l^2}{2eE})|^2} e^{-\pi\omega}\right].
\end{eqnarray*}
To obtain the transmission coefficient we have to compute $|c_1|^2$:
$$
|c_1|^2=\alpha^2 \frac{2\omega} {eE} e^{\pi\frac{\mu_l^2}{2eE}}\frac{\left(\cosh(\pi(\omega-\frac{\mu_l^2}{2eE}))-e^{-\pi\omega}\cosh(\pi\frac{\mu_l^2}{2eE})\right)^2}{\sinh(\pi\omega)\cosh(\pi(\omega-\frac{\mu_l^2}{2eE}))}.
$$
Finally, for the coefficients $|T_l|^2$ and $|R_l|^2$ we obtain:
\begin{eqnarray*}
|T_l|^2&=&
-e^{-\pi\frac{\mu_l^2}{2eE}}\frac{\sinh(\pi\omega)}{\cosh(\pi(\omega-\frac{\mu_l^2}{2eE}))}\\
|R_l|^2&=&e^{-\pi\omega}\frac{\cosh(\pi\frac{\mu_l^2}{2eE})}{\cosh(\pi(\omega-\frac{\mu_l^2}{2eE}))}.
\end{eqnarray*}
Observe that $|R_l|^2-|T_l|^2=1$, as expected for bosons.

As in the Dirac case, the level-crossing region, assuming $eE>0$, is determined by $\omega<0$. Pair production is expected to happen only in this region, thus, for $eE>0$, we must calculate (cf. \ref{wl-sum}):
$$
W_l=\frac 12\sum_\omega\log(1+|T_l(\omega)|^2),
$$
for $\omega<0$.
We have $\sum_{\omega} \mapsto \frac{{\cal T}}{2\pi} \int d\omega$ (cf.
 \cite{bmpp-spindel,gabriel-spindel}), where ${\cal T}$ stays for a finite time interval. An easy computation shows that:
$$
\log(1+|T_i|^2)=\log(|R_i|^2)=\log(1+e^{-\pi\frac{\mu_l^2}{eE}})-\log(1+ e^{2\oom-\pi\frac{\mu_l^2}{eE}}),
$$
and we have to evaluate the integral:
$
\int_{-\infty}^0 d\omega
\log \left( 1 + e^{ 2\pi \omega -  \frac{\pi \mu_l^2}{\ee} } \right)
= -\frac{1}{2\pi} {\mathrm{Li}}_2 \left(
-e^{ -  \frac{\pi \mu_l^2}{\ee} } \right)
$.
In strict analogy with \cite{belcadalla} we obtain:
\beq \label{wk-ucI}
W_l = \frac 12 \frac{{\cal T}}{2\pi} \left[\left(\int_{-\infty}^0 d\omega \right)
\log \left(1 + e^{- \frac{\pi \mu_l^2}{\ee}} \right)
+ \frac{1}{2\pi} {\mathrm{Li}}_2 \left(
-e^{-  \frac{\pi \mu_l^2}{\ee} } \right)\right].
\eeq
The factor $\frac{{\cal T}}{2\pi} \left(\int_{-\infty}^0 d\omega \right)$ amounts to a degeneracy factor  and the same geometric considerations done in \cite{belcadalla} allow us to evaluate it following  \cite{bmpp-spindel}. The degeneracy factor for the scalar case is the same as in the Dirac case and its value is $\frac{{\cal T}}{2\pi} \left(\int_{-\infty}^0 d\omega \right)=eES/2\pi$, with $S=TL$ where $T$ and $L$ are the sizes of the space time box over which $E$ is non vanishing. This value is exactly the same as the one obtaining in the $\zeta$-function approach. The final result (\ref{wk-ucI}) for the imaginary part of the effective action coincides with the result (\ref{effzeta}) we will find using the zeta function approach. It is worth mentioning that the above background implements the physical model
analyzed in \cite{gabriel-spindel}, apart for the fact that in \cite{gabriel-spindel} one deals with
a 2D model and a further parameter $a$ appears (which in our case is equal to 1). The fact that
all our geometries allow a Kaluza-Klein reduction (compare the discussion in \cite{belcadalla})
explains why a correspondence with a 2D model is found: the only substantial difference
is represented in our case by the presence of an effective mass which is given by
$\mu_l^2 = \mu^2 + 2\Lambda l(l+1)$ replacing the mass $\mu^2$ of the aforementioned
2D model.

\subsection{$\zeta$-function approach}

Also for this background  we analyze pair-production with the zeta function method. This technique confirms the results obtained with the transmission coefficients approach.
The Euclidean Klein-Gordon (KG) operator on ultracold I is:
\beq
KG = -\frac{1}{\chi^2} \partial_{\tau}^2 -\frac{1}{\chi} \partial_{\chi} (\chi \partial_{\chi})
-2\Lambda \nabla^2_{\Omega} + \mu^2 + 2 i \ee \frac 12 \partial_{\tau}+ (\ee)^2 \frac 14 \chi^2.
\eeq
In the eigenvalue equation $KG \phi = \lambda \phi$ we put:
\beq
\phi = e^{-i\omega \tau} Y_{lm}(\Omega) \psi(\chi),
\eeq
which leads to variable separation, where $Y_{lm}(\Omega)$ are the usual spherical harmonics appearing in every problems with spherical symmetry. Then we obtain:
\beq
-\frac{1}{\chi} \partial_{\chi} (\chi \partial_{\chi} \psi)+\left[ \mu^2_l + \frac{1}{\chi^2}
\left( \omega + \ee \frac 12 \chi^2 \right)^2 \right] \psi = \lambda \psi.
\eeq
We also introduce $t = \frac 12 \chi^2$, and then we obtain:
\beq
(t\partial_t)^2 \psi + \left[ \frac 12 \left( \lambda -\mu^2_l \right) t -\frac 14
( \omega + \ee t)^2 \right] \psi =0.
\eeq
By choosing
\beq
\psi = e^{ -\frac 12 \ee t } t^{-\frac 12 \omega} g(t),
\eeq
and introducing $z= \ee t$, we obtain the confluent hypergeometric equation:
\beq
z \partial_z^2 g + (1-\omega-z) \partial_z g - \frac 12 \left(1-\frac{\lambda -\mu^2_l}{\ee} \right) g =0.
\eeq
We require that solutions $\psi$ belong to $L^2 [(0,\infty),\frac{dz}{z}]$ (the measure is inherited
from the one of the usual scalar product for scalar particles). It is easy to realize that
this requires to consider different solutions for $\omega<0$ and for $\omega>0$. Let us first consider:
\beq
g (t) = \Phi \left( \frac 12 \left(1-\frac{\lambda -\mu^2_l}{\ee} \right), 1-\omega, \ee t \right).
\eeq
We need the quantization condition:
\beq
\frac 12 \left(1-\frac{\lambda -\mu^2_l}{\ee} \right) = -n,
\eeq
with $n\in {\mathbb N}$, and then:
\beq
\lambda_{n,l} = (2 n +1 ) \ee + \mu^2_l.
\eeq
We have $\psi\in L^2 [(0,\infty),\frac{dz}{z}]$ iff $\omega < 0$.\\
The solution:
\beq
g (t) = t^{\omega} \Phi \left( \frac 12 \left(1-\frac{\lambda -\mu^2_l}{\ee} \right) + \omega,
1+\omega, \ee t \right)
\eeq
requires a further quantization condition:
\beq
\frac 12 \left(1-\frac{\lambda -\mu^2_l}{\ee} \right) + \omega = -n,
\eeq
and then
\beq
\lambda_{n,l,\omega} = (2 n +1 ) \ee + 2\ee \omega + \mu^2_l,
\eeq
and we can conclude that $\psi\in L^2 [(0,\infty),dz]$ iff $\omega > 0$.\\
We obtain that the heat kernel
\beq
K(s) = \sum_l (2 l + 1) k_l(s)
\eeq
receives different contributions from different ranges for $\omega$. In
particular, for $\omega <0$ there is, as in the ultracold II case, a degeneracy
in $\omega$ to be determined, being $\lambda_{n,l}$ independent of $\omega$ in
that region. We get
\beq
k_l (s) = D e^{ -\mu_l^2 s }  e^{ -\ee s } \frac{1}{1-e^{ -2\ee s } },
\eeq
where formally
\beq
D = \int_{-\infty}^0 d\omega.
\eeq
We determine $D$ as in \cite{blissepf}, by comparing the expansion of $k_l (s)$ as $s\to 0^+$
with the heat kernel expansion. We obtain
\beq
D= \frac{\ee S}{2\pi},
\eeq
where $S$ is the volume of the first 2D factor of the metric.\\
We obtain
\beqnl
\zeta_l (s) &=& \frac{\ee S}{2\pi} (2\ee)^{-s} \zeta_H (\frac 12 (1+\frac{\mu_l^2}{\ee}), s) \cr
&+& \frac{{\cal T}}{2\pi}  (2\ee)^{-s} \frac{1}{s-1} \zeta_H (\frac 12 (1+\frac{\mu_l^2}{\ee}), s-1).
\eeqnl
By rotating $\ee \mapsto i\ee$ and looking for $\Imm \zeta'_l (0)$, we obtain
a first contribution from the $\omega<0$ region
which is easily realized to be the same as in the ultracold II case and a further
contribution from the $\omega>0$ region which is given by
$$
\frac{\pi}{2} \Ree \zeta_H (\frac 12 (1+\frac{\mu_l^2}{i\ee}), -1)+
(\log (2\ee)  -1) \Imm \zeta_H (\frac 12 (1+\frac{\mu_l^2}{i\ee}), -1)
- \Imm \zeta'_H (\frac 12 (1+\frac{\mu_l^2}{i\ee}), -1);
$$
as to the first term we get
\beq
\frac{\pi}{2} \Ree \zeta_H (\frac 12 (1+\frac{\mu_l^2}{i\ee}), -1) =
\frac{1}{8\pi} \left[ -{\mathrm{Li}}_2 ( -e^{-\frac{\pi \mu_l^2}{\ee}} )
-{\mathrm{Li}}_2 ( -e^{\frac{\pi \mu_l^2}{\ee}} ) \right];
\eeq
the second one is zero, whereas the third one is
\beq
\Imm \zeta'_H (\frac 12 (1+\frac{\mu_l^2}{i\ee}), -1) =
\frac{1}{8\pi} \left[ {\mathrm{Li}}_2 ( - e^{-\frac{\pi \mu_l^2}{\ee}} )
-{\mathrm{Li}}_2 ( - e^{\frac{\pi \mu_l^2}{\ee}} ) \right].
\eeq
As a consequence, we obtain
\beq \label{effzeta}
\Imm \zeta'_l (0) =- \frac{\ee S}{2\pi} \frac{1}{2} \log(1+e^{-\frac{\pi \mu_l^2}{\ee}} ) -
\frac{{\cal T}}{2\pi} \frac{1}{4\pi} {\mathrm{Li}}_2 (-e^{-\frac{\pi \mu_l^2}{\ee}} )
\eeq
which leads to a full accord with (\ref{wk-ucI}).

As to thermal effects, in this case we limit ourselves to point out that eqn. (\ref{thermal-eff})
holds, but with a pathological behavior associated with the fact that the chemical potential
$\phi$ is ill-defined unless a spatial cut-off is introduced at $\chi=\chi_0<\infty$. The same
phenomenon affects Dirac particles \cite{belcadalla}.

%%%%%%%%%%%%%%%%%%%%%%%%%%%%%%%%%%%%%%%%%%%%%%%%%%%%%%%%%%%%%%%%%%%%%%%%%%%%%%%%%%%%%%%%%%%

\section{Nariai case}
\label{sec-nariai}

We now consider the more general case, that is the electrically charged Nariai solution.
The manifold is described by the metric \cite{romans,bousso,mann}
\beq
ds^2 = \frac{1}{A} (-\sin^2 (\chi) d \psi^2 + d\chi^2) + \frac{1}{B} (d\theta^2+
\sin^2 (\theta) d\phi^2),
\label{nariai-metric}
\eeq
with $\psi \in \RR, \chi\in (0,\pi)$, and the constants
$B=\frac{1}{2 Q^2}\left( 1-\sqrt{1-12 \frac{Q^2}{L^2}} \right)$, $A=\frac{6}{L^2}-B$
are such that $\frac{A}{B}<1$, and $L^2:= \frac{3}{\Lambda}$. The black hole horizon
occurs at $\chi=\pi$. This manifold differs
from the ultracold cases because it has finite spatial section.
In the Euclidean version,
it corresponds to two spheres characterized by different radii.
One finds the following non-vanishing Christoffel symbols $\Gamma_{01}^0 = \cot (\chi),
\Gamma_{00}^1 = \sin (\chi) \cos (\chi), \Gamma_{33}^2 = -\sin (\theta) \cos (\theta),
\Gamma_{23}^3 = \cot (\theta)$.
For the gauge potential we can choose
$A_{i} = -Q \frac{B}{A} \cos (\chi) \delta_{i}^0$.\\
Also for this more complex case we study pair-production making use of the transmission coefficients and zeta function approach. Again these two different methods give the same results. For the Nariai case the zeta function approach requires some mathematical techniques recently developed in \cite{c-zeta} and their application is strictly analogue to the Dirac case, exhaustively analyzed in \cite{belcadalla}.
\subsection{Transmission coefficient approach}
We perform variable separation and set $\phi(\psi,\chi,\Omega)=e^{-i\omega\psi} Y_{lm}(\Omega)\Psi(\chi)$;
moreover, we define $\mu_l^2=\frac {\mu^2}A +\frac BA l(l+1)$.
We need to find the solution of the Klein-Gordon equation for the variable $\chi$:
\beq
\left[-\frac 1{\sin^2 \chi} (\omega -e Q\frac BA \cos\chi )^2 -\frac 1{\sin \chi }\partial_\chi (\sin \chi \partial_\chi)+\mu^2_l\right]\Psi(\chi)=0.
\eeq
Let us first change variable, $t=-\cos \chi$. Then
\beq
(1-t^2)\Psi''-2t\Psi'+\left[\frac 1{1-t^2} (\omega +e Q\frac BA t )^2 -\mu_l^2  \right]\Psi=0,
\eeq
where the prime is the derivation w.r.t. $t$. Note that this equation is invariant under $\{t\rightarrow -t,\ Q\rightarrow -Q\}$ so that
we can look at the singularity in $t=1$ only and obtain the properties of the singularity in $t=-1$ by $Q\rightarrow -Q$. Now, near $t=1$
$$
0\approx 2(1-t) \Psi''-2\Psi'+\frac 1{2(1-t)} (\omega +e Q\frac BA  )^2
$$
which has solution $\Psi =(1-t)^{\pm \frac i2 (\omega +e Q\frac BA)}$. This suggests to set
\beqnl
&& \Psi(t)=(1-t)^{l_+} (1+t)^{l_-} \Phi(t),\\
&& l_\pm= \frac i2 |\omega \pm e Q\frac BA |
\eeqnl
so that the equation for the function $\Phi$ is
$$
(1-t^2)\Phi''-2(t-l_+ (1-t)+l_- (1+t))\Phi' -[\mu_l^2-\omega^2+l_+ +l_- -(l_+-l_-)^2]\Phi=0.
$$
Let us introduce
$$
E:=Q\frac{B}{A}.
$$
We are interested in the level-crossing region, which is, for $\ee>0$,
$$
-e E \leq \omega \leq e E.
$$
In this region one obtains
$$
l_{\pm} = \frac{i}{2} (e E \pm \omega).
$$
We define also
\begin{equation}
\Delta ={\mu_l^2}+(e E)^2-\frac 14.
\label{delta}
\end{equation}
Note that the sign of $\Delta$ is not ensured to be positive. To be precise, we should
also indicate the dependence of $\Delta$ on $l$, by writing e.g. $\Delta_l$, but, in order to
simplify the notation, we leave implicit this dependence. Note also that, if
$\mu^2+(e E)^2-\frac 14<0$, then for sufficiently high values of $l$ the quantity
$\Delta$ passes from negative to positive values. The sign of $\Delta$ is associated
with a different behavior of the transmission coefficients and then of the imaginary part
of the effective action.
A little consideration allows to draw the conclusion that
the behaviors in the two different regions (positive and negative) are linked each other
by analytic continuation.\\
We first consider the case $\Delta>0$.
The general solution of this equation in the level crossing region is easily found to be
\begin{eqnarray}
&&\!\!\!\!\!\Phi(t)=C_+ F\!\left(i e E+\frac 12+i\sqrt{\Delta},
i e E+\frac 12-i\sqrt{\Delta}; i(e E+\omega)+1; \frac {1-t}2\right)\cr
&&\qquad\!\!\!\!\!\!\!\!\ + C_- F\!\left(i e E+\frac 12+i\sqrt{\Delta},
i e E+\frac 12-i\sqrt{\Delta}; i(e E-\omega)+1; \frac {1+t}2\right),
\end{eqnarray}
where $F(a,b;c;z)$ is the usual hypergeometric function.

We can use the well known relation
\beqnl
F(a,b;c;z)&=&\frac {\Gamma(c)\Gamma(c-a-b)}{\Gamma(c-a)\Gamma(c-b)} F(a,b;1+a+b-c;1-z)\cr
&+&  \frac {\Gamma(c)\Gamma(a+b-c)}{\Gamma(a)\Gamma(b)} (1-z)^{c-a-b}\cr
&&\hphantom{} F(c-a,c-b;1-a-b+c;1-z) \label{relaz}
\eeqnl
to look at the asymptotic behavior of $\Psi$ near the boundaries.

Let us now introduce the coordinates $x=\log \tan \frac \chi2$. In this coordinate $\chi\approx 0,\pi$ become $x\approx -\infty, +\infty$
and setting
\begin{eqnarray}
&& \eta(x):=\Psi(\chi(x)),\\
&& \alpha:=\frac {\Gamma(1+i(e E-\omega))\Gamma(-i(\omega+e E))}{\Gamma(\frac 12 -i\omega -i\sqrt \Delta)\Gamma(\frac 12 -i\omega +i\sqrt \Delta)},\\
&& \beta:=\frac {\Gamma(1+i(e E-\omega))\Gamma(i(\omega+eE))}{\Gamma(\frac 12 +i e E +i\sqrt \Delta)\Gamma(\frac 12 +i eE -i\sqrt \Delta)},\\
&& \alpha':=\frac {\Gamma(1+i(\omega+e E))\Gamma(i(\omega- e E))}{\Gamma(\frac 12 +i\omega -i\sqrt \Delta)\Gamma(\frac 12 +i\omega +i\sqrt \Delta)},\\
&& \beta':=\frac {\Gamma(1+i(\omega+e E))\Gamma(i(-\omega+ e E))}{\Gamma(\frac 12 +i e E +i\sqrt \Delta)\Gamma(\frac 12 +i e E -i\sqrt \Delta)},
\end{eqnarray}
we can write
\begin{eqnarray}
&& \eta(x) \approx e^{i(\omega + e E)x} [C_-+C_+ \alpha] +e^{-i(\omega+e E)x} C_+\beta \qquad\ x\approx +\infty,\\
&& \eta(x) \approx e^{i( e E -\omega)x} [C_++C_- \alpha'] +e^{-i(e E-\omega)x} C_-\beta' \qquad\ x\approx -\infty.
\end{eqnarray}
If we are searching for the transmission of a particle coming from $x\approx -\infty$,
then at $t\rightarrow +\infty$ we must find only the transmitted particle at $x\approx +\infty$, with positive momentum.
With the chosen condition, the positive momentum is $(\omega+e E)$ so that we must set $C_+=0$ and $c_{\rm out}=C_-$ is the coefficient of the outgoing
particle. At $x\approx -\infty$, the coefficient of the ingoing particle is then $c_{\rm in}=C_-\alpha'$, so that the transmission
coefficient is
\beq
\tilde{T_l}=\frac {c_{\rm out}}{c_{\rm in}}=\frac 1{\alpha'}=
\frac
{\Gamma(\frac 12 +i\omega -i\sqrt \Delta)\Gamma(\frac 12 +i\omega +i\sqrt \Delta)}
{\Gamma(1+i(\omega+e E))\Gamma(i(\omega- e E))}.
\eeq
Then, by using known formulas for the Gamma function, one obtains
\beq
|\tilde{T_l}|^2 =
\frac{e E-\omega}{e E+\omega} \frac{\sinh (\pi (\ e E -\omega)) \sinh (\pi (\ e E +\omega))}
{\cosh (\pi (\omega - \sqrt \Delta)) \cosh (\pi (\omega + \sqrt \Delta))}.
\eeq
We used a different notation for $\tilde{T_l}$ because it is not yet the transmission
coefficient such that $|R_l|^2=1+|T_l|^2$. The latter is obtained by noticing that
\beq
|T_l|^2 = -\frac{r}{q}|\tilde{T_l}|^2,
\eeq
where $r:=\omega+\ee$ and $q:=\omega-\ee$. Compare also \cite{manogue}, where a
fine discussion upon the topic of the Klein paradox is given.\\
As a consequence, we find
\beq
|T_l|^2 =
\frac{\sinh (\pi (\ e E -\omega)) \sinh (\pi (\ e E +\omega))}
{\cosh (\pi (\omega - \sqrt \Delta)) \cosh (\pi (\omega + \sqrt \Delta))}.
\eeq
In the limit $e E \gg \omega$ one finds
\beq
|T_l|^2 \sim e^{- 2 \pi (\sqrt \Delta- e E)} =
e^{ - 2 \pi e E (\sqrt{1+\frac{\mu_l^2-\frac 14}{(e E)^2}} -1)},
\eeq
which, apart for the term $-\frac 14$, is the result which can be obtained in the WKB approximation.\\
It is easy to show, in the case $\Delta<0$, that the only change consists in the replacement
$i\sqrt{\Delta} \mapsto \sqrt{|\Delta|}$
in the above formulas for the solution and also for $\alpha,\alpha',\beta,\beta'$.
As a consequence, we find the following result:
\beq
|\tilde{T_l}|^2=\frac
{|\Gamma(\frac 12 +i\omega -\sqrt{|\Delta|})|^2 |\Gamma(\frac 12 +i\omega +\sqrt{|\Delta|})|^2}
{|\Gamma(1+i(\omega+e E))|^2 |\Gamma(i(\omega- e E))|^2};
\eeq
the denominator is the same as in the case $\Delta>0$. As to the numerator one finds
$$
|\Gamma(\frac 12 +i\omega -\sqrt{|\Delta|})|^2 |\Gamma(\frac 12 +i\omega +\sqrt{|\Delta|})|^2 =
\frac{\pi}{\cos (\pi z_1)} \frac{\pi}{\cos (\pi z_2)}
$$
where $z_1:=\sqrt{|\Delta|}+i\omega$ and $z_2:=\sqrt{|\Delta|}-i\omega$ and standard relations
for the Gamma function are used. As a consequence we get
\beq
|\tilde{T_l}|^2 =\frac{e E-\omega}{e E+\omega} 2\frac{\sinh (\pi (\ e E -\omega)) \sinh (\pi (\ e E +\omega))}
{\cosh(2\pi \omega)+\cos (2\pi \sqrt{|\Delta|})}.
\eeq
We need to calculate
\beq
W =\frac 12 \sum_i \log (1+|T_i|^2);
\eeq
we do not perform the sum over $l$, and then
we calculate:
\beq
W_l = (2 l+1) \frac 12 \sum_{\omega} \log (1+|T_l (\omega)|^2).
\eeq
Let us start from the case $\Delta>0$. We have to perform the following integral:
\beq
I:=\int_{-eE}^{eE} d\omega \log (1 +
\frac{\cosh [2\pi eE]-\cosh [2 \pi \omega]}{\cosh [2 \pi \sqrt \Delta]+\cosh [2 \pi \omega]})
\eeq
where the dependence on $l$ is implicit in $\Delta$; the integral can be rewritten as follows:
\beq
I = 2 eE \log (\cosh [2\pi \sqrt \Delta]+\cosh [2 \pi eE])-II,
\eeq
where
\beqnl
II:&=&\int_{-eE}^{eE} d\omega \log (\cosh [2\pi \sqrt \Delta]+\cosh [2 \pi \omega])\cr
&=&\frac{1}{2\pi} \int_{-2\pi eE}^{2\pi eE} dy \log (p+\cosh [y]),
\eeqnl
which is formally the same integral as in the Dirac case.
Then we find
\beqnl
W_l &=& (2 l+1) \frac{{\cal T}}{2\pi} \left( \ee \log \left[ 2 \left( \cosh [2\pi \sqrt \Delta]+ \cosh [2 \pi eE]\right) \right] \right.\cr
&+& \left. \frac{1}{4\pi}
\left[- {\mathrm{Li}}_2 (-e^{-2\pi (\sqrt{\Delta}+eE)} ) +  {\mathrm{Li}}_2 (-e^{2\pi (\sqrt{\Delta}+eE)} )    \right. \right. \cr
&-&\left. \left.
 {\mathrm{Li}}_2 (-e^{2\pi (\sqrt{\Delta}-eE)} )+ {\mathrm{Li}}_2 (-e^{-2\pi (\sqrt{\Delta}-eE)}) \right] \right).
\eeqnl
As to the case $\Delta<0$, it can be obtained by the replacement $\sqrt{\Delta}\mapsto i \sqrt{|\Delta|}$.
In particular, if $\mu^2+(\ee)^2-\frac{1}{4}<0$, one finds
that there exists $l_c$ such that
\beq
W = \sum_{l\leq l_c} \tilde{W}_l + \sum_{l>l_c} W_l,
\eeq
where
\beqnl
\tilde{W}_l &=&\frac{{\cal T}}{2\pi} (2l+1)
\left( \ee \log \left[ 2 \left( \cos [2\pi \sqrt{|\Delta|}]+
\cosh [2 \pi eE] \right) \right] \right. \cr
&+& \left. \frac{1}{4\pi}
\left[- {\mathrm{Li}}_2 (-e^{-2\pi (i \sqrt{|\Delta|}+eE)} )
- {\mathrm{Li}}_2 (-e^{2\pi (i \sqrt{|\Delta|}-eE)} ) \right. \right. \cr
&+&\left. \left.
{\mathrm{Li}}_2 (-e^{2\pi (i \sqrt{|\Delta|}+eE)})
+ {\mathrm{Li}}_2 (-e^{-2\pi (i \sqrt{|\Delta|}-eE)}) \right] \right).
\eeqnl
By taking into account that ${\mathrm{Li}}_2 (\bar{z}) = \overline{{\mathrm{Li}}_2 (z)}$,
it is evident that the latter expression is real.

%%%%%%%%%%%%%%%%%%%%%%%%%%%%%%%%%%%%%%%%%%%%%%%%%%%%%%%%%%%%%%%%%%%%%%%%%%%%%%%%%%%
\subsection{Nariai in the $\zeta-$function approach}

The Euclidean Klein-Gordon operator for the Nariai solution is given by %(\cite{belcaccia-rnds}, eq. 14)
\beq
-\frac{A}{\sin^2\chi}(\partial_{\tau}-i\ee\cos\chi)^2-\frac{A}{\sin\chi}\partial_\chi(\sin\chi\partial_\chi)-B\nabla^2_\Omega+\mu^2\equiv KG^{(E)},
\eeq
where $\tau = i\psi$.
Let us search for the eigenfunctions of this differential operator. We can perform variable separation,
as usual. Note that $B \nabla^2_\Omega$ and $-i\partial_{\tau}$ commute with $KG^{(E)}$ and then one
can restrict the study of its eigenvalue equation to the eigenspaces of the aforementioned
operators, i.e. we can write for its eigenfunctions in these eigenspaces
$f(\tau,\chi,\Omega)=e^{-i\omega \tau} Y_{l,m}(\Omega) g(\chi)$,
where $Y_{l,m}(\Omega)$ are the spherical harmonics. Moreover, note that
\beq
(-B\nabla^2_\Omega+\mu^2)Y_{l,m}=\mu_l^2 Y_{l,m} \qquad \mu_l^2=\frac{\mu^2}{A}+\frac{B}{A}l(l+1).
\eeq
The operator $KG^{(E)}|_{\omega,l,m}$ restricted to the above eigenspaces takes the
following form:
\beq
KG^{(E)}|_{\omega,l,m}=
\left [-\frac{A}{\sin^2\chi}(i\omega-i\ee\cos\chi)^2-
\frac{A}{\sin\chi}\partial_\chi(\sin\chi\partial_\chi)
+A\mu^2_l\right].
\eeq
Let us introduce the new variable $t=-\cos\chi$; then the eigenvalue equation for  $KG^{(E)}|_{\omega,l,m}$
becomes:
\beq \label{iper1}
\left[ -\frac{1}{1-t^2}\left( \omega+\ee t\right)^2 -\mu^2_l+\frac{\lambda}{A}\right ]g+(1-t^2)g''-2t g'=0.
\eeq
To transform this equation into an hypergeometric, we set $g(t)=(1+t)^{l_-}(1-t)^{l_+}\psi(t)$, with
\begin{equation}
l_\pm=\frac{1}{2}\left|\omega\pm \ee \right |.
\end{equation}
This choice ensures that the solutions belong into the Hilbert space $L^2((-1,1))$ for all values
of $\omega\in\rr$.
As a consequence, equation (\ref{iper1}) becomes:
\beqnl
&&(1-t^2)\psi''(t) + \left[ -2t + 2l_+ (1-t) -2l_- (1+t) \right] \psi'(t)\cr
&&+\left[-\omega^2-\mu^2_l+\frac{\lambda}{A}-2l_+l_-+l_+^2-l_++l_-^2-l_-\right]\psi(t)=0.
\eeqnl
The general solution of this equation is
\beq
\psi(t)={}_2F_1(a_+,a_-;2l_-+1;\frac{1+t}{2})+{}_2F_1(a,b;2l_++1;\frac{1-t}{2}),
\eeq
with
\beq
a_{\pm}=\frac{1}{2}+l_++l_- \pm \sqrt{\frac{1}{4}-\omega^2-\mu_l^2+\frac{\lambda}{A}+(l_++l_-)^2+(l_+-l_-)^2}.
\eeq
Note that this solution has a bad behavior in $t=\pm1$. The only possibility to for it to lie in $L^2((-1,1))$ is that $a_-\in -\mathbb{N}$, that is
\beq
\frac{1}{2}+l_+ + l_- - \sqrt{\frac{1}{4}-\frac{\mu^2_l}{A}+\frac{\lambda}{A}+(\ee)^2}=-n,\qquad n \in \NN.
\eeq
Indeed, the spectrum is discrete with eigenvalues
\beq
\frac{\lambda_{n,l,\omega}}{A}=\left[n+\frac{1}{2} +(l_+ + l_-)\right]^2- \frac{1}{4}+\frac{\mu^2_l}{A}-(\ee)^2,
\eeq
which are degenerate in the azimuthal quantum number $m$. \\
It follows that the heat kernel for the operator $KG^{(E)}$ is $k(s)=\sum_{l} k_l(s)$, where
\begin{eqnarray}
k_l(s) &=\trace e^{-sKG^{(E)}} \\ \nonumber
&= \sum_\omega\sum_n (2l+1) e^{  -sA\left(n+\frac{1}{2}+(l_++l_-)\right)^2- \frac{1}{4}+\frac{\mu^2_l}{A}-(\ee)^2}.
\end{eqnarray}
Actually the sum over $\omega$ is an integral due to the continuity of the $-i\partial_t$ spectrum.
It is convenient to split such integration into two parts that are the interval $-\ee <\omega<\ee $ and its complement in $\mathbb{R}$.
This is because inside the interval the eigenvalues $\lambda_{n,\omega,l}$ do not depend on $\omega$. In this way, we get
\begin{eqnarray*}
k_l(s) &= \frac{{\cal T}}{2\pi} \left( 2(2l+1)\int_{\ee }^\infty\sum_n e^{-sA\left[\left(n+\frac{1}{2}+\omega\right)^2+\frac{\mu_l^2}{A}-(\ee)^2
-\frac{1}{4}\right]} d\omega \right.\\
&\left. + (2l+1)\left(\ee\right) \sum_n e^{-sA\left[\left(n+\frac{1}{2}+\ee \right)^2+\frac{\mu_l^2}{A}-(\ee)^2
-\frac{1}{4}\right]} \right).
\end{eqnarray*}
The spectral Riemann $\zeta$-function associated to the Klein-Gordon operator with kernel $k_l(t)$ is then
\begin{eqnarray*}
\zeta_l(s) &= \frac{1}{\Gamma(s)}\int t^{s-1}k_l(t)dt \\
&=\frac{{\cal T}}{2\pi} \left( 2(2l+1)\int_{\ee }^\infty\sum_n \frac{d\omega}{A^s\left[\left(n+\frac{1}{2}+\omega\right)^2+\frac{\mu_l^2}{A}-(\ee)^2
-\frac{1}{4}\right]^s} \right.\\
&\left. + (2l+1)\left(\ee\right) \sum_n \frac{1}{A^s\left[\left(n+\frac{1}{2}+\ee \right)^2+\frac{\mu_l^2}{A}-(\ee)^2
-\frac{1}{4}\right]^s}\right).
\end{eqnarray*}
An analogous computation of one performed in \cite{belcadalla,c-zeta} leads to the following expression
for the imaginary part of the effective action:
\begin{eqnarray}\label{action2}
W_l %=-\Imm (\frac{d}{ds}|_{s=0}\zeta^L_l(s))
&=&\frac{{\cal T}}{2\pi} (2l+1)\left(
\ee\log \left[\cosh [\pi(\sqrt \Delta -\ee)] \cosh [\pi(\sqrt \Delta +\ee)]\right]
 \right. \cr
&& \left. +2 \ee\log 2 +\frac{1}{4\pi} \left[- {\mathrm{Li}}_2 (-e^{-2\pi (\sqrt{\Delta}+eE)})+ {\mathrm{Li}}_2
(-e^{2\pi (\sqrt{\Delta}+eE)})
\right. \right.\cr
&& \left. \left.  - {\mathrm{Li}}_2 (-e^{2\pi (\sqrt{\Delta}-eE)})+ {\mathrm{Li}}_2
(-e^{-2\pi (\sqrt{\Delta}-eE)}) \right] \vphantom{\frac{{\cal T}}{2\pi}} \right),
\end{eqnarray}
which coincides with  the one obtained using the transmission coefficient approach.
The same considerations as for the aforementioned approach in the case $\Delta<0$
apply in the zeta-function approach.

\subsection{Thermal effects}

We find (for definiteness we choose $\Delta>0$)
\beqnl
<\bar{N}_l^{out}>_{\beta_h} &=&
\frac{\sinh (\pi (\ e E -\omega)) \sinh (\pi (\ e E +\omega))}
{\cosh (\pi (\omega - \sqrt \Delta)) \cosh (\pi (\omega + \sqrt \Delta))}\cr
&&\times \frac 12 \left( \coth [\pi (\omega-\varphi^+)]+
\coth [\pi (|\omega|+\varphi^-)] \right),
\label{inst-nariai}
\eeqnl
with $\varphi^+ = e (A_0 |_{\pi}-A_0 |_{0})= 2 \ee=\varphi^-$.
We recall that in terms of physical
(dimensionful) variables, by taking into account that
$T_h = \frac{\hbar c \sqrt{A}}{2\pi k_b}$, and that $\omega_{phys}=\sqrt{A} \omega$,
in such a way that $\beta_{phys} \omega_{phys}=2 \pi \omega$.

\section{Conclusions}
\label{conclusion}

Our analysis for the scalar case has confirmed the main features we obtained in \cite{belcadalla}:
exact calculations have been performed, both in the transmission coefficient approach and in the
zeta-function approach. The latter is more involved but it also provides us much more information
with respect to the former, indeed the complete one loop effective action (and not only its
imaginary part) can be obtained by using the zeta-function, as known. Differences with the
Dirac case are both of general character (indefinite scalar product spaces vs. Hilbert spaces)
and in particular characteristics of the cases we analyzed: in the ultracold I case a bad behavior
at $x=+\infty$, which does not occur for the Dirac case, forced us to refer to fluxes in order to
compute the transmission coefficient; moreover, in the Nariai case, the coefficient $\Delta$
is not ensured to be positive-definite (whereas it is positive definite in the Dirac case),
and then one is forced to consider both cases. It is worth mentioning that
this aspect is not new, because an analogous
problem occurs in the well-known case of the so-called Sauter potential; nevertheless, a discussion
of that problem for the Sauter potential is often missing (cf. e.g. \cite{niki}, where
the case associated with our $\Delta<0$ is considered, and e.g. the results in \cite{manogue} and in
\cite{kimpage}(first paper),
for Sauter-like potentials, where the opposite case is given).
Thermal effects, as in the Dirac case, have been shown
to affect the discharge phenomenon, with a key-role in the pair-creation phenomenon
still to be assigned to the transmission coefficient.\\

%%%%%%%%%%%%%%%%%%%%%%%%%%%%%%%%%%%%%%%%%%%%%%%%%%%%%%%%%%%%%%%%%%%%%%%%%%%%%%%%%%%%%%%%%%%%%%%%%%

\end{document}